\documentclass[twocolumn, times]{aastex631}
\usepackage{comment}

\shorttitle{Acoustic Wave Energy Fluxes and Radiative Losses in the Solar Chromosphere}
\shortauthors{da Silva Santos et al.}
\newcommand{\jcr}[1]{\textcolor{red}{#1}}

\newcommand{\uat}[2]{\href{http://astrothesaurus.org/uat/#2}{#1 (#2)}}

\DeclareRobustCommand{\ion}[2]{\textup{#1\,\textsc{\lowercase{#2}}}}

\graphicspath{{./}{figures/}}

\begin{document}

\title{Constraints on Acoustic Wave Energy Fluxes and Radiative Losses in the Solar Chromosphere from Non-LTE Inversions}

\correspondingauthor{Jo\~{a}o Santos}
\email{jdasilvasantos@nso.edu}


\author[0000-0002-3009-295X]{J. M. da Silva Santos}
\affiliation{National Solar Observatory, 3665 Discovery Drive, Boulder, CO 80303, USA}

\author[0000-0003-0583-0516]{M. Molnar}
\affiliation{High Altitude Observatory, NCAR, P.O. Box 3000, Boulder, CO 80307, USA}


\author[0000-0002-0189-5550]{I. Mili\'{c}}
\affiliation{Leibniz Institute for Solar Physics (KIS), Sch\"{o}nekstr. 6, 79104 Freiburg, Germany}
\affiliation{Faculty of Mathematics, University of Belgrade, Studentski Trg 16, 11000 Belgrade, Serbia}

\author[0000-0001-5850-3119]{M. Rempel}
\affiliation{High Altitude Observatory, NCAR, P.O. Box 3000, Boulder, CO 80307, USA}

\author[0000-0001-8016-0001]{K. Reardon}
\affiliation{National Solar Observatory, 3665 Discovery Drive, Boulder, CO 80303, USA}

\author[0000-0002-4640-5658]{J. de la Cruz Rodríguez}
\affiliation{Institute for Solar Physics, Dept. of Astronomy, Stockholm University, AlbaNova University Centre, 106 91, Stockholm, Sweden}



\begin{abstract}
Accurately assessing the balance between acoustic wave energy fluxes and radiative losses is critical for understanding how the solar chromosphere is thermally regulated. We investigate the energy balance in the chromosphere by comparing deposited acoustic flux and radiative losses under quiet and active solar conditions using non-local thermodynamic equilibrium (NLTE) inversions with the Stockholm Inversion Code (STiC). To achieve this, we utilize spectroscopic observations from the Interferometric BIdimensional Spectrometer (IBIS) in the \ion{Na}{I} 5896\,\AA~and \ion{Ca}{II} 8542\,\AA~lines and from the Interface Region Imaging Spectrograph (IRIS) in the \ion{Mg}{II} h and k lines to self-consistently derive spatially resolved velocity power spectra and cooling rates across different heights in the atmosphere. Additionally, we use snapshots of a three-dimensional radiative-magnetohydrodynamics simulation to investigate the systematic effects of the inversion approach, particularly the attenuation effect on the velocity power spectra and the determination of the cooling rates. The results indicate that inversions potentially underestimate acoustic fluxes at all chromospheric heights while slightly overestimating the radiative losses when fitting these spectral lines. However, even after accounting for these biases, the ratio of acoustic flux to radiative losses remains below unity in most observed regions, particularly in the higher layers of the chromosphere. We also observe a correlation between the magnetic field inclination in the photosphere and radiative losses in the low chromosphere in plage, which is evidence that the field topology plays a role in the chromospheric losses. 
\end{abstract}


\section{Introduction} \label{sec:intro}

The quiescent chromosphere radiates an average of 2/0.3/4\,$\rm kW\,m^{-2}$ (low/upper/whole chromosphere) of energy \citep{1977ARA&A..15..363W}, mostly through emission in the \ion{Ca}{II} lines \citep[e.g.,][]{1981ApJS...45..635V}. For a long time, wave energy dissipation has been considered one of the main processes heating the solar chromosphere, providing the energy that is lost via radiation.  Waves are generated by convective motions in the photosphere and propagate upwards into the chromosphere and corona. In an ideal and homogeneous magnetized plasma, the solutions of the magnetohydrodynamic (MHD) equations result in fast, slow, and intermediate (or Alfvén) wave modes. As the waves propagate, they interact with the magnetic field, undergoing reflection, refraction, and mode conversion. Ultimately, these interactions lead to wave dissipation and the heating of the surrounding plasma \citep[][]{2015SSRv..190..103J}. In particular, the velocity amplitudes of acoustic waves (when the phase velocity equals the sound speed) increase with height due to the negative density gradient in the chromosphere, causing them to steepen into shocks \citep{1946NW.....33..118B,1948ApJ...107....1S}. 

Evidence supporting acoustic wave heating has been provided, for example, by observations and modeling of the characteristic Doppler and intensity oscillations in the \ion{Ca}{II} H, K, and 8542\,\AA~lines in the weakly magnetized areas of the quiet-Sun \citep[QS, e.g.,][]{1991SoPh..134...15R, 1993ApJ...414..345L,1997ApJ...481..500C,2009A&A...494..269V}. Non-local thermodynamic equilibrium (NLTE) inversions of high-resolution observations have confirmed that the properties of bright grains in the internetwork are consistent with propagating shocks, showing upflow speeds of up to $\sim$\,6\,$\rm km\,s^{-1}$ and temperature enhancements of up to $\sim$\,4,500\,K \citep{2022A&A...668A.153M}. Observations at millimeter wavelengths have shown continuum brightness temperature enhancements of the same order when accounting for the difference in spatial resolution relative to visible wavelengths  \citep{2020A&A...644A.152E,2021A&A...656A..68E}. Furthermore, ultraviolet (UV) observations have linked internetwork bright grains to propagating shocks reaching the transition region in some instances \citep{2015ApJ...803...44M,2018MNRAS.479.5512K}. Even in the absence of shocks, dissipation of the wave energy is possible through ion-neutral collisions \citep{2016ApJ...819L..11S,2020A&A...635A..28W}. 

Acoustic shocks also partly contribute to enhancing emissions in plage regions \citep{2007A&A...466.1131R, 2016ApJ...826...49S, 2021A&A...648A..28A}, where a magnetic-field-dependent component is also expected \citep[e.g.,][]{2018A&A...619A...5B}. In fact, the observed radiative losses appear to be a superposition of different effects operating on different timescales \citep{2022A&A...664A...8M}. These effects potentially include resistive heating at the edges of expanding flux tubes \citep{2013ApJ...764L..11D}, Alfvén wave turbulence \citep{2011ApJ...736....3V}, current-inducing vortex flows \citep{2021A&A...645A...3Y}, and magnetic braiding \citep{2024NatAs.tmp...66B}, among others, which have been considered in plages. 
 
However, the exact contribution of acoustic waves to the energy balance of the quiescent chromosphere has remained controversial in recent years; some authors have found approximately enough acoustic flux to sustain the canonical radiative losses \citep{2009A&A...508..941B,2010A&A...522A..31B,2020A&A...642A..52A, 2021A&A...648A..28A}, whereas other measurements fall short by up to one order of magnitude \citep{2005Natur.435..919F,2006ApJ...646..579F,2007PASJ...59S.663C,2009A&A...507..453B,2012A&A...544A..46B,2016ApJ...826...49S,2021ApJ...920..125M}. Small-scale magnetic reconnection between emerging magnetic fields and the ambient field in the QS has been considered a potential supplementary energy source; yet, recent observations indicate that it occurs too infrequently \citep{2018ApJ...857...48G,2024ApJ...964..175G}. 

The discrepancies mentioned above may be attributed to the limiting effect of spatial resolution in some observations \citep{2007ApJ...657L..57C,2007ApJ...671.2154K}, the use of different spectral diagnostics sensitive to different layers of the atmosphere, and underlying methodological assumptions, such as the mass density and attenuation coefficients of the atmosphere, which are usually taken from spatially averaged, semi-empirical models or numerical simulations external to the data and thus might not be adequately constrained \citep{2023arXiv230204253M}. 

This paper addresses the aforementioned shortcomings by presenting self-consistent, spatially resolved heating rates and wave fluxes inferred from NLTE inversions of spectroscopic observations of the internetwork and plage acquired by the Interface Region Imaging Spectrograph \citep[IRIS,][]{2014SoPh..289.2733D} conjointly with the Interferometric BIdimensional Spectrograph \citep[IBIS,][]{2006SoPh..236..415C} at the Dunn Solar Telescope \citep[DST,][]{1991AdSpR..11e.139D}. We also analyze synthetic observables calculated from a radiative MHD simulation to quantify the systematic errors of our approach. 

\section{Data}
\label{sec:Observations}

\begin{figure*}
	\includegraphics[width=\textwidth]{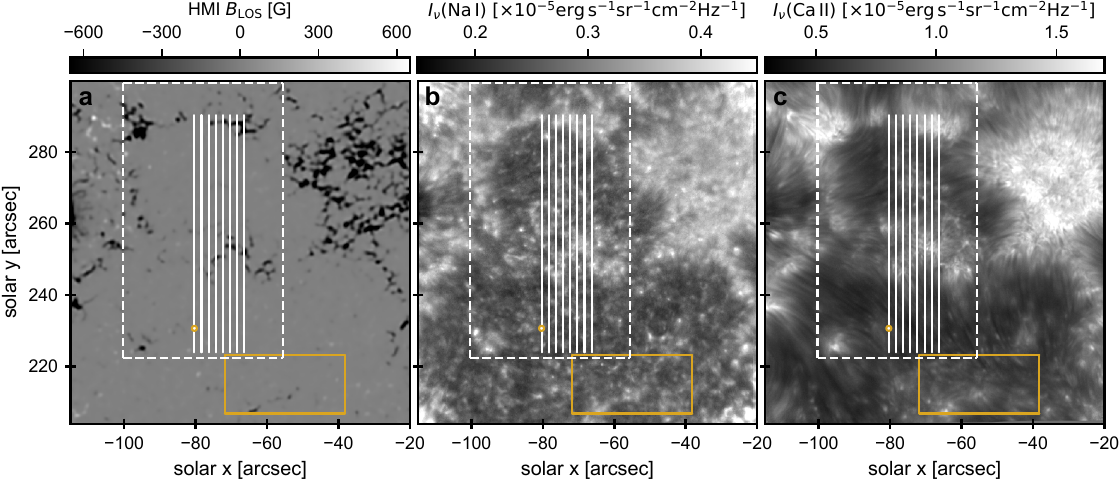}
	\caption{Overview of the target as observed by SDO/HMI and IBIS at 15:54 UT on 23 April 2017. Panel (a): HMI LOS magnetogram; the colorbar range is capped for display purposes. Panel (b): Intensity in the core of the \ion{Na}{I}\,D$_{1}$ line. Panel (c): Intensity in the core of the \ion{Ca}{II}\, 8542\AA~line. The vertical solid lines show the eight IRIS slit positions. The dashed box outlines the Hinode/SP FOV. The solid yellow box delimits the "quiet patch" selected for inversions. The yellow circle marks the location referred to in Fig.\,\ref{fig:inv2}. }
	\label{fig:FOV_overview}
\end{figure*}

\subsection{Observational data}

We used spectroscopic and spectropolarimetric observations of the leading edge of NOAA active region (AR) 12651, with the center of the field-of-view (FOV) located at heliocentric coordinates [-77\arcsec, 254\arcsec] at $\mu$=0.96, where $\mu$ is the cosine of the heliocentric angle. These observations captured a plage region and the surrounding internetwork. The observing campaign included partial temporal and spatial overlaps from different observatories, including the Atacama Large Millimeter/Submillimeter Array \citep[ALMA,][]{2009IEEEP..97.1463W}, the spectropolarimeter of the Solar Optical Telescope \citep[SOT/SP,][]{2008SoPh..249..167T} onboard Hinode, DST/IBIS, and IRIS. Recent studies have analyzed the co-temporal IBIS/ALMA datasets to explore the correlation between the width of the $\rm H\alpha$ line and the 3\,mm continuum brightness \citep{2019ApJ...881...99M}, determine acoustic fluxes using 1D modeling of the different spectral diagnostics \citep{2021ApJ...920..125M}, and infer the temperature stratification from NLTE inversions \citep{2022ApJ...933..244H}.

In this paper, we analyze the co-temporal IBIS, IRIS, and Hinode observations of the same region. The ALMA Band 6 data were not included in the analysis, as \citet{2022ApJ...933..244H} highlighted the challenges in modeling these observations alongside the visible spectral lines. These challenges may arise from the low resolution of the ALMA maps, calibration uncertainties, and/or the need to account for time-dependent hydrogen ionization, which is not feasible with the available inversion codes \citep{2022FrASS...9.7878W}. 


IBIS was run in intensity mode, prioritizing high temporal cadence in the \ion{Na}{I} D$_1$\,5896\,\AA~and \ion{Ca}{II} 8542\,\AA~lines, hereafter $\lambda5896$ and $\lambda8542$. The spectral sampling (range) is 0.015 (0.8)\,\AA~and 0.05 (2.1)\,\AA~in $\lambda5896$ and $\lambda8542$, respectively, with a total line scan cadence of approximately 10\,s. According to the Nyquist theorem, this allows us to resolve temporal frequencies up to 50\,mHz. The spatial pixel scale is $0.096^{\arcsec}$ per pixel. A more detailed explanation of the IBIS observing mode and data reduction can be found in \citet{2021ApJ...920..125M} and \citet{2022ApJ...933..244H}. 

IRIS observed part of the IBIS FOV using an 8-step coarse raster centered on a small plage region at the FOV center, between 15:25--19:00 UT. The cadence of the observations is 26\,s with a single exposure of 2\,s at each raster step. We only analyzed the \ion{Mg}{II} h an k wavelength window as the signal-to-noise ratio of the far-UV lines was inadequate for inversion analysis (\S\,\ref{section:methods_nlteinversions}). The IRIS spectra were calibrated to absolute flux units using standard radiometric calibration. We coaligned the IBIS and IRIS datasets by determining the offsets between the $\lambda8542$ core images and the IRIS slitjaw images in the 2800\,\AA~passband. We rebinned the common FOV to a plate scale of $0.16^{\arcsec}$, that is, slightly degrading the IBIS scans to the IRIS resolution. 

Hinode/SOT/SP scanned a region 45$^{\arcsec}$ across, encompassing the IRIS FOV with a cadence of about 10\,min. Here, we used the Level 2 data products, consisting of the magnetic field components in the photosphere obtained through Milne-Eddington inversions of the \ion{Fe}{I} 6301/6302\,\AA~lines \citep{2013SoPh..283..601L}. The slit step is $\sim$\,0.297\arcsec~and the pixel scale is $\sim$\,0.319\arcsec~along the slit. 

Figure~\ref{fig:FOV_overview} shows an overview of the target as observed by the different telescopes. The closest-in-time photospheric line-of-sight (LOS) magnetogram provided by the Helioseismic and Magnetic Imager \citep[HMI,][]{2012SoPh..275..207S} is displayed for context. The magnetogram has been deconvolved and super-resolved (to $0.3^{\arcsec}$/px) using the \texttt{Enhance} machine learning code\footnote{\url{https://github.com/cdiazbas/enhance}} \citep{2018A&A...614A...5D}. 
The white vertical lines indicate the different IRIS slit positions, and the yellow box defines the "quiet patch" in our analysis. This quiet patch, for which we only have IBIS data, was selected to be as close as possible to quiescent conditions, away from strong field concentrations and opaque chromospheric fibrils as seen in the core of $\lambda8542$. 

\subsection{Simulated data}
\label{section:methods_simulation}

We used a 3D radiative-MHD simulation of the QS performed with the \texttt{MURaM} code 
\citep{2005A&A...429..335V, 2017ApJ...834...10R} to investigate how well inversions can recover velocity power spectra and the radiative energy losses in the chromosphere. This simulation was derived from the case 'O16bM' in \citet{2014ApJ...789..132R} after extending the simulation domain further up into the corona. The simulation has a horizontal extent of $24.6$\,Mm and a vertical extent of $16.4$\,Mm with a uniform grid spacing of 32\,km. The average continuum $\tau=1$ level is located about 6.2\,Mm above the bottom boundary. A small-scale magnetic dynamo maintains a mixed polarity field with quiet Sun level field strength in the convection zone part of the simulation. At the top boundary, we impose a temperature of 1.1\,MK to aid the formation of a transition region, as a fully self-maintained QS corona is difficult to achieve in a domain with the relatively small horizontal extent chosen here. The upper boundary is open, which is implemented through a symmetric condition on mass density and vertical flow velocity. While this minimizes reflections of the outgoing disturbances, it does lead to downflows at the top boundary that reached a mean value of around $20\rm\,km\,s^{-1}$ after 15 minutes of simulated time. We used 100 snapshots obtained at a cadence of 9\,s. To manage the computational load, we used only a subset of 400 pixels per time step, evenly distributed throughout the simulation domain, providing sufficient statistics. The chromosphere is treated with gray radiative transfer (RT) and ionization balance in LTE. Although the simulated conditions are not designed to precisely replicate the observational targets, they provide a reasonable environment to investigate the methodological approach described in \S\,\ref{section:methods_radlosses} and \S\,\ref{section:methods_acoustic_flux}.


\section{Methods}

\subsection{NLTE inversions}
\label{section:methods_nlteinversions}

We performed NLTE inversions of the IBIS and IRIS spectra to obtain the stratification of temperature, density, line-of-sight (LOS) velocity, and radiative cooling rates in the atmosphere. To that end, we used the STockholm Inversion Code \citep[\texttt{STiC},][]{2016ApJ...830L..30D,2019A&A...623A..74D}, which is a Message Passing Interface (MPI)-parallel code built upon a modified version of the Rybicki \& Hummer \citep[RH,][]{2001ApJ...557..389U} code. 
\texttt{STiC} employs an optimization algorithm that minimizes the merit (or cost) function, $\chi^{2}$, defined as
\begin{equation}
    \chi^2(\textbf{p}) = \frac{1}{N_{\lambda}}\sum^{N_{\lambda}}_{i=1} \left[\frac{y_{i}-f(\lambda_{i};\textbf{p})}{\sigma_{i}}\right]^2 + \vphantom{\sum^{\rm N}_{i=1}}\sum^{N_{\rm p}}_{k=1}\alpha_{k} r_{k}(\textbf{p})^2, 
    \label{eq:reg_chisq}
\end{equation}
\noindent where the first term is the mean sum of squared residuals between the observed data, $y_{\rm i}$, and the model, $f(\lambda_{i};\textbf{p})$, acting on a set of parameters, $\textbf{p}$, weighted by a weighting function, $\sigma_{\rm i}$, for all observed wavelength points, $\lambda_{\rm i}$, and the second term is the regularization term controlled by the weights, $\alpha_k$, that adjust the magnitude of the regularization functions, $r_k$, for each parameter. Regularization helps in convergence; for example, we penalize the second derivative of temperature to impose some degree of smoothness in the model \citep{2019A&A...623A..74D}. 
To investigate the differences in the inverted atmospheres obtained from different spectral diagnostics, we ran the inversions in two modes: IBIS data alone using the $\lambda$5896 and $\lambda$8542 lines, and IBIS+IRIS including the \ion{Mg}{II} h, k, and subordinate triplet lines within the IRIS passband when possible. 

We solved the statistical equilibrium equation for the atomic population densities using a 12-level atom model for \ion{Na}{I}, a 6-level atom model for \ion{Ca}{II}, and a 11-level atom model for \ion{Mg}{II}, all including a continuum level from higher ionization species. Other atoms and molecules of abundant species were treated in LTE. We investigated the impact of including the \ion{H}{I} atom in the NLTE with electron densities corrected via charge conservation \citep{2019A&A...623A..74D} and found that this had only a small impact (few percent) on the derived temperatures. 
Therefore, in the interest of saving up to a factor of five in computational cost for a few percent change in the cooling rates (\S\,\ref{section:methods_radlosses}), which is well within the uncertainties of the inversions (\S\,\ref{section:methods_uncertainties}), \ion{H}{I} was treated in LTE. 

Similarly to \citet{2022ApJ...933..244H}, the inversions were performed in column mass scale ($\xi = p_{\rm gas}/g$, where $g$ is the solar surface gravity and $p_{\rm gas}$ is the gas pressure) instead of the traditional optical depth scale to better resolve the chromosphere. In the IBIS mode, we used five nodes in LOS velocity, $v_{\rm LOS}$, and microturbulence, $v_{\rm turb}$, and eight nodes in temperature, $T$. In the IBIS+IRIS mode, we used up to eleven nodes in $T$ and eight nodes in $v_{\rm LOS}$ and $v_{\rm turb}$. 

The drawback of this space-time-resolved inversion approach is its computational cost, making it virtually unfeasible when solving the statistical equilibrium equations for several atoms simultaneously, even for a moderate-sized high-resolution FOV. To address this, we accelerated the inversions using a neural network (NN), which, after adequate training, provides close-to-optimal initialization for regular NLTE inversions with \texttt{STiC}, as detailed in the Appendix~\ref{section:appendix}.

\subsection{Radiative energy losses}
\label{section:methods_radlosses}

\begin{figure*}[t]
    \centering
    \includegraphics[width=\linewidth]{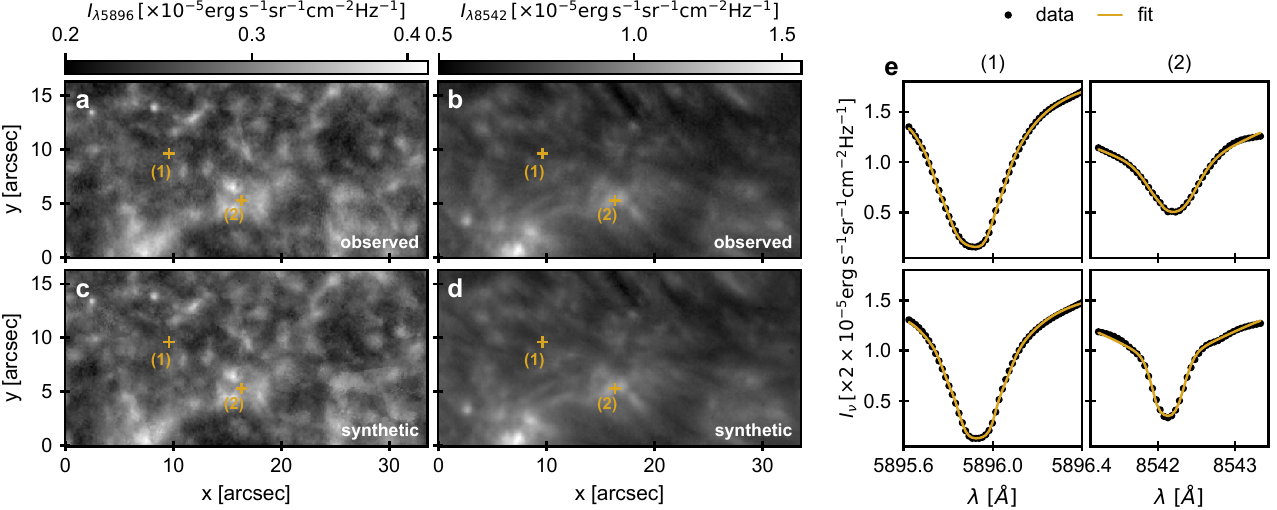}
    \caption{Inversions of the quiet patch. Panels (a-b): Observed intensities in the core of $\lambda$5896 and $\lambda$8542. Panel (c-d): Corresponding synthetic intensities. Panels (e): Example observed spectra (dots) and best fit (lines) at the two locations marked in the left panels. } \label{fig:inv1} 
\end{figure*}

We computed the radiative energy losses (or net cooling rates) in \ion{Ca}{II} (H, K, and infrared triplet lines), \ion{Mg}{II} (h, k, and UV triplet lines), and \ion{H}{I} (Ly-$\alpha$) using the inverted atmospheres. In a bound-bound transition between upper level $j$ and lower level $i$, the net radiative cooling rate is given by 
\begin{equation}
    Q_{ji}= h\nu(n_{j}R_{ji} - n_{i}R_{ij}), \label{eq:losses}
\end{equation}
\noindent where $h$ is the Planck constant, $n_{j}$ and $n_{i}$ are the number densities of the levels, and $R_{ji}$ and $R_{ij}$ are the downward and upward radiative rates, respectively. More explicitly,  Eq.\,\ref{eq:losses} can be written as \citep[e.g.,][]{2002ApJ...565.1312U}
\begin{eqnarray}
    Q_{ji} & = & \frac{h\nu}{4\pi}\oint d\Omega\int d\nu  \bigg[ n_{j}A_{ji}\psi_{ij}(\nu,\textbf{n}) \nonumber \\ 
    & - & B_{ij} \left( n_{i}-\frac{g_{i}}{g_{j}}\rho_{ij}(\nu, \textbf{n})n_{j}\right)\phi_{ij}(\nu, \textbf{n}) I(\nu, \bf{n})\bigg],
\end{eqnarray}
\noindent where $A_{ji}$ and $B_{ij}$ are Einstein coefficients \textcolor{violet} for spontaneous emission and absorption respectively, $g_{i}$ and $g_{j}$ are the statistical weights of the levels, $\rho_{ij}$ is the ratio of the emission, $\psi_{ij}$, and absorption profiles, $\phi_{ij}$, and $I(\nu, \bf{n})$ is the intensity for a given frequency $\nu$ and direction $\bf{n}$. \jcr{}After obtaining model atmospheres from the \texttt{STiC} inversions (\S \ref{section:methods_nlteinversions}), we ran the code once more in synthesis mode to obtain the populations and rates to evaluate Eq.\,\ref{eq:losses}. 
The Ly-$\alpha$ losses were included for completeness, although their contribution to the total losses is only a couple percent in the lower chromosphere and only becomes significant in the temperature regime above $\sim$\,10,000\,K \citep[see also][]{1981ApJS...45..635V}, which we do not investigate in this paper. 

To investigate the variation of the net cooling rates with height, we computed the total losses in two height bins in the chromosphere for practical purposes: from the temperature minimum up until the $\tau$\,$=$\,1 layer of the core of $\lambda8542$ (referred to as the low-chromosphere, $Q_{\rm low}$), and between the latter and the $\tau$\,$=$\,1 layer of the core of $\lambda2796$ (referred to as mid-chromosphere, $Q_{\rm mid}$). This accounts for the fact that different atmospheres have different height scales, making thresholds in geometrical height/optical depth or temperature unreliable. 
Our definitions are comparable to what was loosely described in the early literature as the low-chromosphere -- the layers immediately after the temperature minimum up until the $\sim$\,6,000\,K temperature knee in the classical models, the upper-chromosphere -- the layers above 10,000\,K, and the mid-chromosphere being everything in between the other two \citep[e.g.,][]{1977ARA&A..15..363W}. 

We also investigated the impact of LTE and NLTE \ion{H}{I} populations and electron densities on the losses for a sample of pixels in the quiet patch and plage region. We concluded that treating \ion{H}{I} in NLTE has only a minor effect in the lower chromosphere but becomes more significant in the mid and upper chromosphere due to the increasing contribution from Ly-$\alpha$. 
Even so, the deviations compared to the LTE case are generally within the estimated (1$\sigma$) uncertainty. We find that the average difference in $Q_{\rm low}$($Q_{\rm mid}$) derived from models assuming LTE versus NLTE \ion{H}{I} populations is $\sim$\,4\%(20\%).

\subsection{Inversion uncertainties}
\label{section:methods_uncertainties}

We estimated the uncertainties of the inversions following a Monte Carlo approach \citep[e.g.,][]{1992nrca.book.....P}. This is a computationally intensive task, where the fitting is repeated multiple times with different realizations of the noise. Consequently, we cannot provide standard errors for the parameters for every pixel in the FOV. Instead, we focused on selected pixels to gauge the typical uncertainties in the inversion parameters and radiative losses. 

At any given location in the FOV, we generated 100 different line profiles by adding a random noise component to the synthetic spectra resulting from a good fit to the observations, taking into account the measured noise variance of the signals and the absolute flux calibration uncertainties. For the IBIS lines, the flux calibration is based on the high-resolution solar atlas published in \citet{1999SoPh..184..421N}, with a photometric uncertainty of less than 0.5\%. The IRIS flux calibration depends on routine, accurate measurements of the spectrograph effective areas, involving cross-calibration with SOLSTICE with an absolute accuracy of 5\%~\citep{2018SoPh..293..149W}. Furthermore, the synthetic spectra were inverted using different randomly generated atmospheres as initial guesses to assess the impact of the starting solution. Finally, we computed the standard deviations of the parameters at different column masses across the independent inversion runs. 

In the region observed both by DST/IBIS and IRIS, the uncertainties in temperature in the IBIS+IRIS mode typically increase from less than 50\,K ($\lesssim$1\%) in the photosphere up to several hundred kelvin ($\lesssim$16\%) in the upper chromosphere. Uncertainties in LOS velocity are similar in both inversion modes, increasing from $\lesssim$\,0.5\,$\rm km\,s^{-1}$ in the photosphere to $\lesssim$\,1\,$\rm km \,s^{-1}$ where the $\lambda$2796 core forms. The propagating uncertainties in the total radiative losses (as defined in \S\,\ref{section:methods_radlosses}) are on the order of $\sim$\,14\% in the low chromosphere in the IBIS mode and $\sim$\,5\%(18\%) in the low(mid) chromosphere in the IBIS+IRIS mode. 


\subsection{Acoustic flux}
\label{section:methods_acoustic_flux}

An estimate of the acoustic flux, $F_{ac}$, for vertically propagating waves in the solar atmosphere can be obtained with the 
following equation \citep[e.g.,][]{1974soch.book.....B}:
\begin{equation}
    F_{\rm ac} = \rho \sum_{\nu} \frac{\left < v_{\rm obs}^2 (\nu) \right > }{\cal{T}(\nu)} v_{\rm gr}(\nu),
    \label{eqn:acoustic_flux}
\end{equation}

\noindent where $\rho$ is the plasma mass density at the formation height of the spectral line, $\left < v_{\rm obs}^2 (\nu) \right >$ is the amount of oscillatory power in the $\nu$
frequency bin, $\cal{T}$ is the attenuation coefficient (or transfer function), and $v_{\rm gr}(\nu)$ is the group velocity given by $v_{\rm gr}(\nu)=c_{\rm s}\sqrt{1-\nu_{\rm ac}/\nu}$, where $\nu_{\rm ac}$ is the acoustic cut-off frequency, and $c_{\rm s}$ is the sound speed. Both $\rho$ and $c_{\rm s}$ were averaged in time for evaluating Eq.\,\ref{eqn:acoustic_flux} at each pixel.

The velocity power spectral densities (PSDs) were computed using the inferred $v_{\rm LOS}$ at different heights. The sound speed was computed from the thermodynamic variables in the models for each pixel, resulting in a spatial variation of the acoustic cutoff frequency, given by $\nu_{\rm ac}$\,$=$\,$\gamma g / 4\pi c_{\rm s}$, in the range $\sim$\,4.5$-$6\,mHz, while the nominal value is at 5.2\,mHz for the canonical value of $c_{\rm s}$\,$=$\,7$\rm\,km\,s^{-1}$ for an adiabatic index $\gamma$\,$=$\,$5/3$, and gravitational acceleration $g$\,$=$\,$274\rm\,m\,s^{-1}$ \citep[e.g.,][]{1977A&A....55..239B}.
In the presence of magnetic field, the acoustic cutoff is modified via the component of gravity along
the field via $\nu_{\rm B}$\,$=$\,$ \nu_{\rm ac}\cos(\theta)$, where $\theta$ is the inclination angle, becoming even more spatially dependent \citep[e.g.,][]{2006ApJ...648L.151J, 2011A&A...534A..65S}. 

We considered including  magnetic field inclination values obtained by HMI in the analysis of the quiet patch. However, we found the values to be mostly unreliable due to low signal-to-noise ratio in Stokes Q and U in the weakly magnetized regions \citep[see also][]{2020A&A...642A..52A}. Hinode/SOT/SP did not scan the quiet patch region, so we lack reliable magnetic field measurements there. However, we did include Hinode measurements in the analysis of the small plage region at the center of the IBIS FOV (Fig.~\ref{fig:FOV_overview}).

The attenuation coefficient is meant to compensate for the fact that the amplitudes of the detected Doppler velocities are attenuated from the real vertical velocity of the solar plasma \citep[e.g.,][]{1980A&A....84...96M} due to a combination of RT effects and the wavelength of the waves being similar to the width of the spectral line's formation region. The attenuation is typically stronger at higher frequencies. Previous studies have either set this coefficient to unity \citep[e.g.,][]{2016ApJ...826...49S,2020A&A...642A..52A} or have taken different approaches to estimating it, including using perturbative approaches
on semi-empirical models \citep[e.g.,][]{1980A&A....84...99S, 2009A&A...508..941B} or using dynamic models \citep[e.g.,][]{2006ApJ...646..579F, 2023arXiv230204253M}. Here, we estimated the total attenuated acoustic flux at different heights based on \texttt{STiC} inversions of synthetic spectra from a 3D \texttt{MURaM} simulation, as explained in \S\,\ref{section:muram_results}.


\section{Results}

\subsection{Inversions of IBIS spectra: the quiet patch}
\label{section:Results:InversionsOfIBIS}

Figure\,\ref{fig:inv1} shows an overview of the inversion results for the first time stamp of the quiet IBIS patch shown in Fig.\,\ref{fig:FOV_overview}. 
In general, the images in the core of both lines show similar structures except in the upper right corner of the FOV, where relatively short, intermittent fibrils partly obscure the reverse granulation pattern in the $\lambda8542$ core but not in $\lambda5896$.
Comparison between panels (a)-(c) and (b)-(d) shows that we generally obtained excellent fits to the $\lambda$5896 and $\lambda$8542 intensities. Panels (e) confirm the quality of the fits for two example spectra. We obtained qualitatively similar results for the remainder of the time series.

\subsubsection{Formation heights of the sodium and calcium lines}
\label{section:Results:formationNaCa}

\begin{figure}
    \centering
    \includegraphics[width=\linewidth]{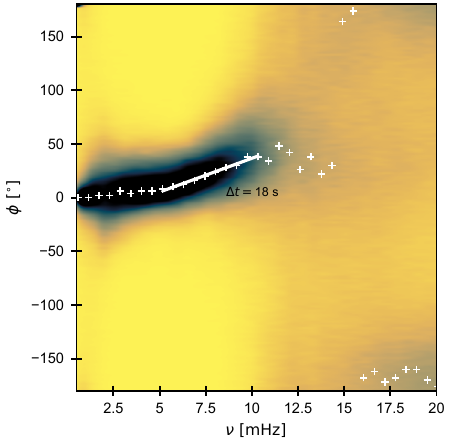}
    \caption{Phase difference between the velocities in $\lambda$5896 and $\lambda$8542. Two dimensional histogram of phase values (inverse color, arbitrary scaling) obtained from the inversion models in the quiet patch as a function of frequency. The markers show the maximum value in each frequency bin. The solid white line shows a linear fit to the underlying points corresponding to the time delay, $\Delta t$.} 
    \label{fig:phase2d}
\end{figure}

\begin{figure*}[t]
    \centering
    \includegraphics[width=\linewidth]{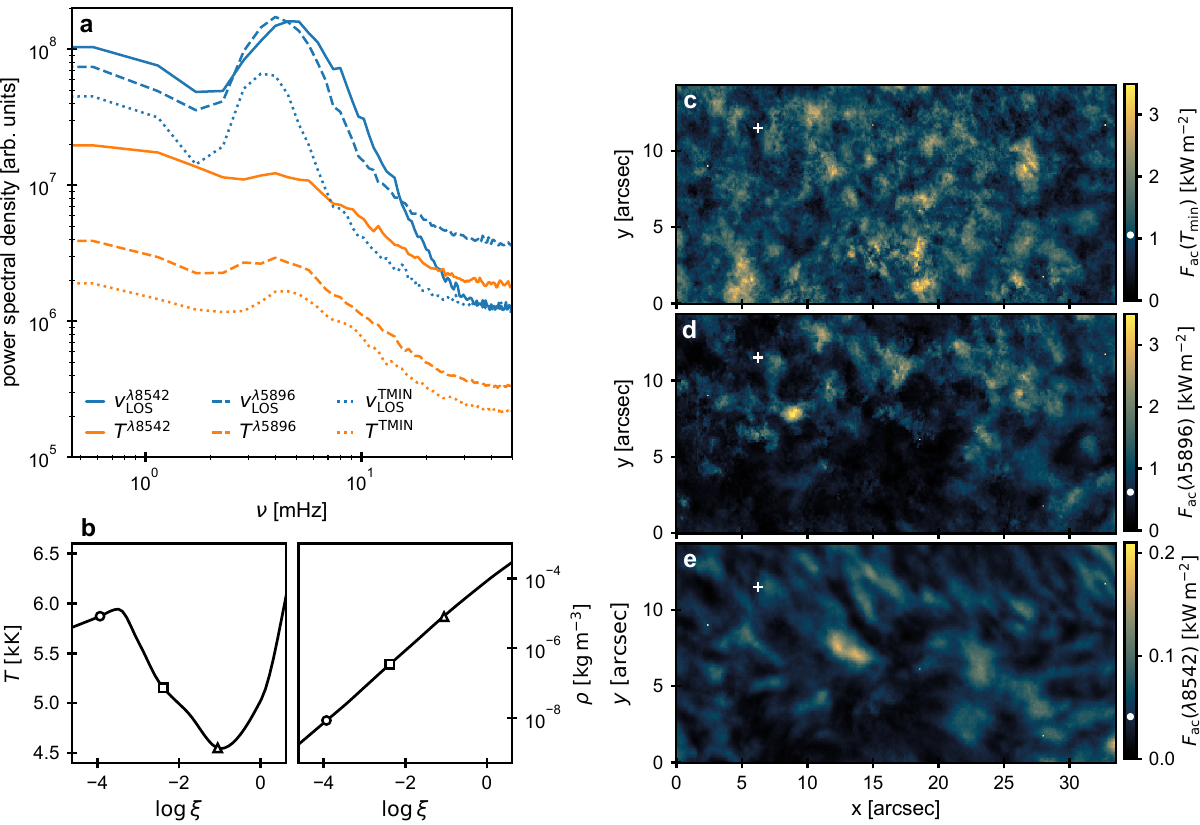}
    \caption{Temperature and velocity oscillations in the quiet patch. Panel (a): Power spectral densities of inverted LOS velocities and temperatures at the temperature minimum (dotted), the core of $\lambda$5896 (dashed), and the core of $\lambda$8542 (solid). Panels (b): Temperature and mass density as a function of logarithmic column mass at a selected location shown by the white cross on the right panels; the three black markers show the column mass values of the temperature minimum (triangle), the core of $\lambda$5896 (square), and the core of $\lambda$8542 (circle). Panels (c-e): Acoustic fluxes at the three considered heights. The white dots on the colorbars indicate median values. }
    \label{fig:inv1_PS} 
\end{figure*}

We estimated the formation height separation between $\lambda$5896 and $\lambda$8542 by calculating the wavelength-dependent opacities from the inversion models and tracking the $\tau_{\lambda}=1$ layer for both lines at each pixel. We find that, on average, the core of $\lambda5896$ and $\lambda8542$ form at $\log \xi$\,$\sim$\,$-2.4$ and $\log \xi$\,$\sim$\,$-3.7$, respectively, in our models (Fig.~\ref{fig:inv1_PS}\textcolor{xlinkcolor}{b}).
Further assuming hydrostatic equilibrium, the corresponding geometrical heights (above the optical depth unity layer of the 5000\,\AA~continuum) are $z$\,$\sim$\,$830(\pm 60)$\,km for $\lambda5896$ and $z$\,$\sim$\,$1200(\pm 40)$\,km for $\lambda8542$, implying an average separation between the two line cores of approximately 370\,km. 

For reference, the average mass density at the formation height of $\lambda8542$ is $\rho$\,$\sim$\,$2(\pm 1)$\,$\times$\,$10^{-8}\rm\, kg\,m^{-3}$. This is a factor of four higher than the value adopted by \citet{2021ApJ...920..125M} based on \texttt{RADYN} simulations. We also find the sound speed to be approximately $c_{\rm s}$\,$\sim$\,$7.5(\pm 0.4)\rm\,km\,s^{-1}$ (mean$\pm$standard deviation) at the formation height of $\lambda8542$, ranging between $c_{\rm s}$\,$\sim$\,6$-$9$\rm\,km\,s^{-1}$ around the FOV. At the temperature minimum height, $c_{\rm s}$ is within 6$-$8\,$\rm km\,s^{-1}$.

Alternatively, the formation height separation can also be directly estimated from the phase (or time delay) between the velocity oscillations measured at the formation heights of $\lambda5896$ and $\lambda8542$. Figure~\ref{fig:phase2d} presents the two-dimensional histogram of phase values, $\phi$, as a function of frequency for all pixels in the quiet patch. The phase angle was obtained from the cross power spectral density function of the LOS velocities at the core of both lines. 
We took into account the fact that the lines are not observed simultaneously but in sequence, with a relative delay of 3.3\,s. The white crosses indicate the maximum of the phase angle distribution at different frequencies. The diagram shows a region of nearly flat phase below 5\,mHz followed by a region with a positive slope up until $\sim$11\,mHz, which is the expected behavior of evanescent waves below the acoustic cutoff frequency and propagating waves at higher frequencies \citep[e.g.,][]{1982ApJ...253..367L}. The phase delay, $\Delta t$\,$=$\,18\,s, was derived from the slope of the fitted line as $\Delta t$\,$=$\,$(2\pi)^{-1}\,d\phi/d\nu$. Assuming the sound waves propagate with $v_{\rm ph}$\,$=$\,$c_{s}=$\,$7.5$\,$\rm km\,s^{-1}$, we derive an height separation between the line cores of only $\sim$\,130\,km, which is at odds with the stark difference in the core intensity in both lines (Fig.~\ref{fig:inv1}). 
Conversely, assuming the mean separation of 370\,km obtained from the inversions, the estimated phase speed is $\sim$\,21$\rm \,km\,s^{-1}$.

\subsubsection{Temperature and velocity power spectra}
\label{section:Results:IBISpower}

\begin{figure}
    \centering
    \includegraphics[width=\linewidth]{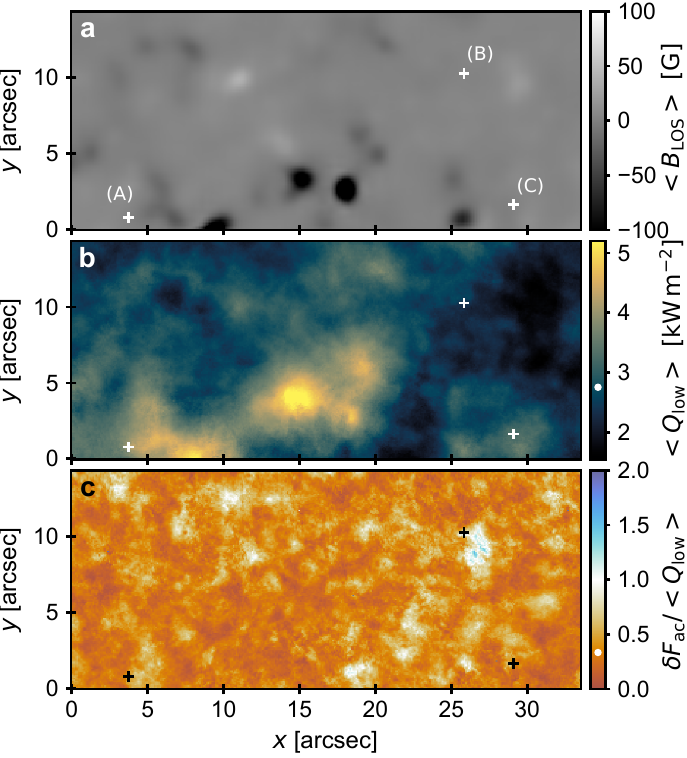}
    \caption{Radiative losses and deposited acoustic flux in the low chromosphere. Panel (a): Time-averaged, deconvolved HMI LOS magnetogram; the range is capped at $\pm$\,100\,G for display purposes. Panel (b): Time-averaged radiative losses in the low chromosphere. Panel (c): Ratio between the deposited acoustic flux and the radiative losses. The crosses show the three locations displayed in Fig.\,\ref{fig:inv1_series}. The white dots on the colorbars show median values.}
    \label{fig:deltaFac} 
\end{figure}

Figure~\ref{fig:inv1_PS} shows the PSDs obtained from the inversions of the quiet patch. The power spectra were computed at three distinct heights: the height of the temperature minimum at the base of the chromosphere, and the heights where the optical depth is unity in the core of $\lambda$5896 and $\lambda$8542, typically between $\log \xi$\,$\sim$\,$[-4, -2]$ (or $\log \tau_{5000}$\,$\sim$\,$[-5, -3]$), where the mass density has dropped by more than two orders of magnitude with height (Fig. \ref{fig:inv1_PS}\textcolor{xlinkcolor}{b}). 
The density drop largely dominates the decrease in the acoustic flux with height. The peak of the velocity power spectra shifts to higher frequencies with height (Fig.\,\ref{fig:inv1_PS}\textcolor{xlinkcolor}{a}), which is the well-known transition from 5-min in the photosphere to 3-min oscillations in the chromosphere. However, the same trend is not observed in temperature, which shows relatively weaker power enhancements around those periodicities. This could still be indicative of heating by acoustic shocks.

As expected, the acoustic flux (Eq.\,\ref{eqn:acoustic_flux}) decreases significantly from the temperature minimum to the formation height of $\lambda8542$. We capped the integration limit in frequency at 20\,mHz above which we reach the noise floor \citep{2021ApJ...920..125M}. 
The noise level was subtracted from each power spectra prior to frequency integration. The mean($\pm$ standard deviation) values at the three heights are 1.1\,($\pm 0.5$), 0.6\,($\pm 0.4$), and 0.04\,($\pm 0.02$) $\rm kW\,m^{-2}$ from the bottom to the top. These values were obtained under the assumption that the attenuation coefficient is unity (Eq.\,\ref{eqn:acoustic_flux}). However, numerical experiments using the \texttt{MURaM} simulation show that they could be underestimated from about 30\% at the temperature minimum region to 80\% at the formation height of $\lambda8542$, as presented in \S\,\ref{section:muram_results}.

\subsubsection{Deposited acoustic flux vs radiative losses}
\label{section:Results:IBISdepflux}

\begin{figure*}
    \centering
    \includegraphics[width=\linewidth]{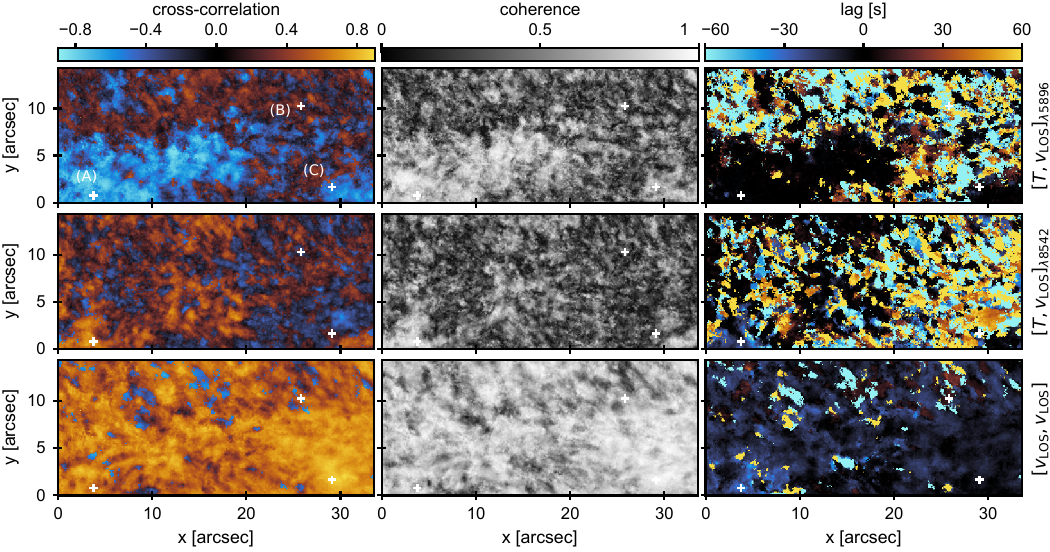}
    \caption{Lead-lag analysis of inverted temperature and velocity at two height ranges in the quiet patch. From the left to the right: cross-correlation coefficient, coherence, and lag between temperature and velocity at the formation height of $\lambda5896$ (top) and  $\lambda8542$ (middle), and between velocities in the two lines at different heights (bottom). The white crosses show the three locations displayed in Fig.\,\ref{fig:inv1_series}.}
    \label{fig:corrs}
\end{figure*}

\begin{figure*}
    \centering
    \includegraphics[width=\linewidth]{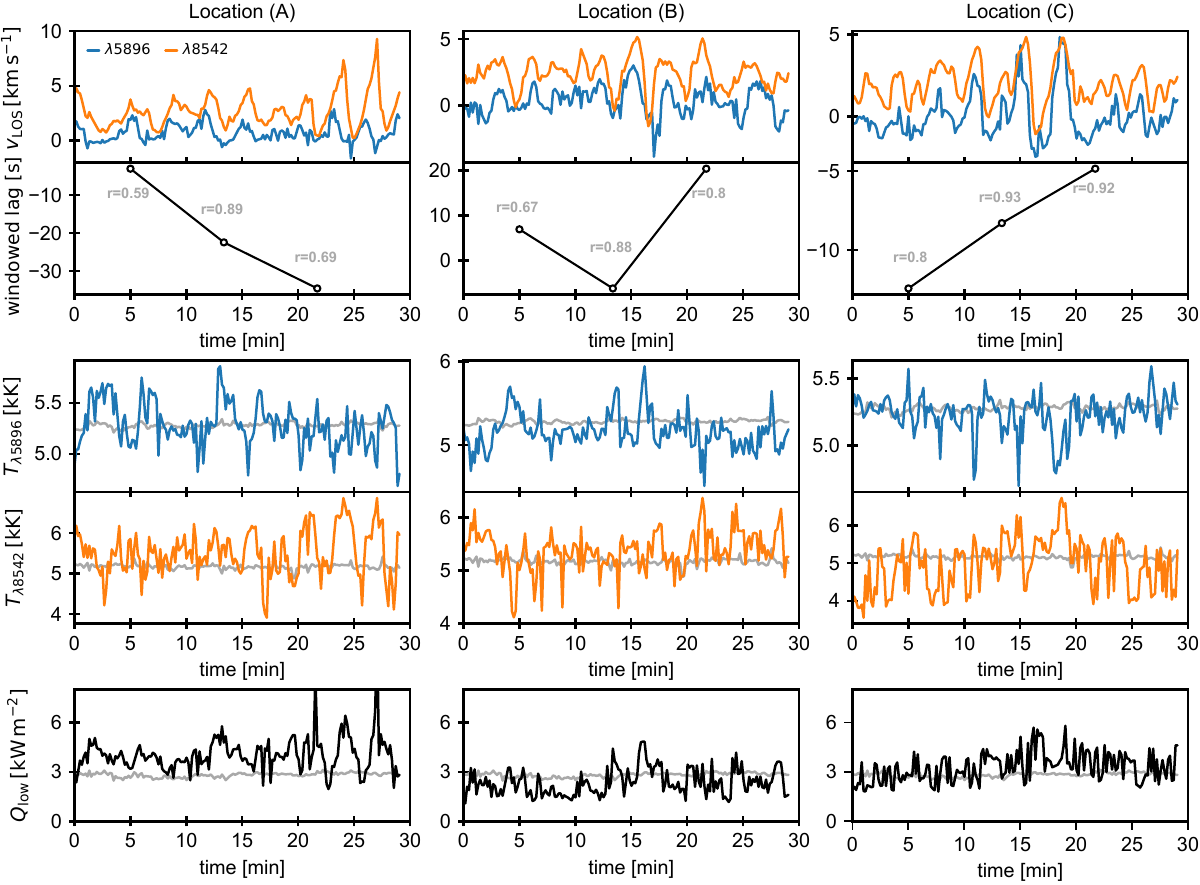}
    \caption{ Time variation of velocity, temperature, and total cooling rates in the low chromosphere in the quiet patch. From the top to the bottom: LOS velocities, windowed correlation lag (with correlation coefficients, $r$), temperature, and integrated radiative losses as a function of time for the pixel locations (A), (B), and (C) marked in Fig.~\ref{fig:deltaFac} and Fig.~\ref{fig:corrs}. The gray lines show (spatially) averaged quantities in the FOV.} \label{fig:inv1_series}
\end{figure*}

Figure \ref{fig:deltaFac} displays the time-averaged total radiative losses in the low chromosphere and the ratio of the deposited acoustic flux to these losses. We calculated the deposited acoustic flux in the low chromosphere as the difference between the acoustic flux at the temperature minimum and the acoustic flux at the formation height of the core of $\lambda8542$. A time-averaged HMI LOS magnetogram is also shown for context.

The radiative losses exhibit significant spatial structure, with a median value of $\sim$\,2.8\,$\rm kW\,m^{-2}$, which is slightly higher than the canonical value of 2\,$\rm kW\,m^{-2}$; however, differences in the integration method and target characteristics must be considered. The enhanced losses are spatially associated with the  magnetic field concentrations detected by HMI, but the correlation between $|B_{\rm LOS}|$ and $Q_{\rm low}$ is weak ($r$\,$=$\,$0.32$), indicating a nonlinear relationship between them. Some of the highest values ($\gtrsim$\,5\,$\rm kW\,m^{-2}$) occur at locations where the height of the temperature minimum may not be well constrained, indicating a flat temperature distribution across a range of column masses; this could introduce a bias into the determination of the integration range. 

The deposited acoustic fluxes are lower than the radiative losses in the majority of the FOV. The median is $\delta F_{\rm ac}$\,$\sim$\,0.9\,$\rm kW\,m^{-2}$, corresponding to $\sim$\,33\% of the losses in the low chromosphere in the quiet patch. 
We also find a few small patches where the acoustic flux is on the order of or even slightly higher than the estimated cooling rates.

\subsubsection{Time-domain analysis}
\label{section:Results:IBIStimeseries}

\begin{figure*}
    \centering
    \includegraphics[width=\linewidth]{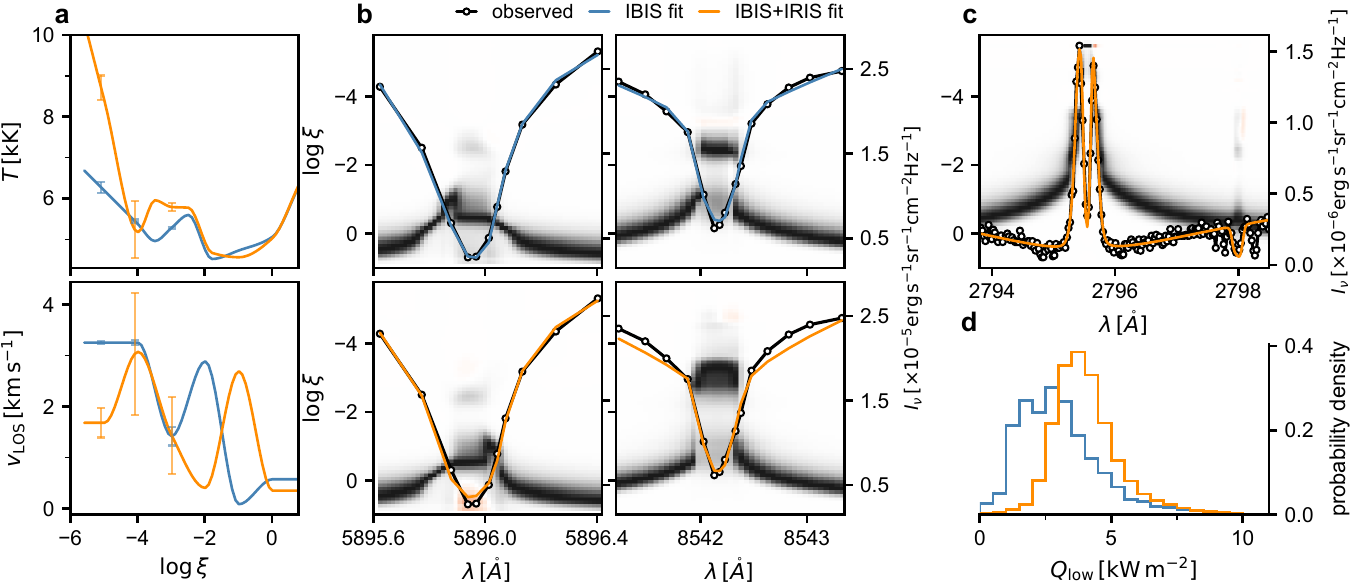}
    \caption{Atmospheric stratification and radiative losses inferred from the IBIS and IRIS spectra. In all panels, the blue and orange lines correspond to the IBIS and IBIS+IRIS fits, respectively. Panels (a): Temperature (top) and LOS velocity (bottom) at the the location marked by the circle in Fig.\,\ref{fig:FOV_overview}; the error bars show the uncertainties in the chromosphere. Panels (b): The foreground shows observed and best-fit intensities in the $\lambda5896$ (left) and $\lambda$8542 (right) lines, and the background shows normalized temperature response functions as a function of column mass and wavelength for the corresponding models shown in panels (a). Panel (c): Analogous to panels (b) but for a wavelength window around the $\lambda2796$ line. Panel (d): Histograms of integrated radiative losses in the low chromosphere for all IRIS slit positions.} 
    \label{fig:inv2}
\end{figure*}

Figure \ref{fig:corrs} presents the time series analysis for velocities and temperature. The cross-correlation coefficient (left panels) was normalized to be directly comparable to the Pearson correlation coefficient. The time lag (or shift) corresponding to the peak of the cross-correlation function is shown in the rightmost panels. Complementing this, the coherence between two signals was calculated as the squared magnitude of their cross-spectral density divided by the product of their individual PSDs, quantifying the consistency of their phase relationship at a given frequency, shown here at 5\,mHz.  These metrics were calculated for the velocities in different layers and between the velocities and temperatures at the same layers at two different heights.

In general, the $\lambda5896$ and $\lambda8542$ lines exhibit highly coherent and positively correlated velocity oscillations with a relatively small lag ($\lesssim$\,20 s) between them (bottom panels), which is consistent with the phase diagram fit (Fig.~\ref{fig:phase2d}). 
However, we noticed that the phase shifts often significantly change over time at some locations (see below).
In some patches, the lag is exactly zero, suggesting standing waves, possibly originating from wave reflections from higher layers of the chromosphere \citep{2021A&A...645L..12F,2024arXiv240619789K}. There are a few isolated patches where the correlation shows weak negative values ($r$\,$\sim$\,$[-0.6, -0.4]$), which are associated with high lags and are likely not realistic. 
The temperature-velocity relationships appear more complex, with a higher prevalence of incoherent oscillations, especially at the formation height of $\lambda8542$ (middle panels). This complexity is partly due to the higher uncertainty of the temperatures compared to the velocities. However, at the core height of $\lambda5896$, we identified an extended patch showing a clear anti-correlation between temperature and velocity with a lag close to zero (top panels). This feature is absent in $\lambda8542$. 
Regions with high (absolute) lag values generally correspond to areas of low coherence and weak linear correlation between temperature and velocity in both lines. 

To further investigate the variation of the velocity correlations and phase delays on shorter time scales, we employed windowed cross-correlation. This technique was not applied to temperatures, as they are more prone to noise. The windowed cross-correlation method divides the signals into overlapping segments. Each segment is multiplied by a Hann window to minimize edge effects, and the cross-correlation is computed within each window. We divided the signals into three $\sim$\,10\,min windows with an overlap of $\sim$\,2\,min. Narrower windows and/or smaller overlaps provide less robust results; larger windows and overlaps are less sensitive to trend changes. All cross-correlation coefficients presented here are statistically significant ($p<0.001$).

Figure~\ref{fig:inv1_series} shows the temporal variation of velocities, temperatures, and radiative losses for three different locations, (A), (B), and (C), around the FOV, as marked in Fig.~\ref{fig:deltaFac} and Fig.~\ref{fig:corrs}, illustrating the range of lead-lag behaviors in the $\lambda5896$ and $\lambda8542$ parameters. 
In particular, we observe steadily increasing (A) or decreasing (C) lags over time and even phase reversals (B) where $\lambda8542$ leads or lags for part of the time interval. About half of the quiet patch shows at least one such reversal. Locations showing overall weak linear correlations and low coherence (Fig.~\ref{fig:corrs}) often coincide with those exhibiting these reversals.

We also observe a range of velocity-temperature relationships, from strong positive or strong negative correlations to a complete lack of correlation. For instance, location (A) shows an $[v_{\rm LOS}, T]$ anti-correlation for $\lambda5896$, although only for part of the time interval. In contrast, for $\lambda$8542, $[v_{\rm LOS}, T]$ are uncorrelated for the first $\sim$\,15\,min, after which a clear positive correlation develops. The strong velocity (up to $\delta v_{\rm LOS}$\,$\sim$\,$6.8\,\rm~km\,s^{-1}$) and temperature (up to $\delta T$\,$\sim$\,$1400\rm~K$) oscillations in $\lambda8542$ between $\sim$\,$[20, 30]$\,min are also associated with a clear, synchronous variation of the radiative losses, which more than double during that time interval. 

Using the measured velocity perturbation of $\delta v_{\rm LOS}$\,$\sim$\,6.8\,$\rm km\,s^{-1}$ above the mean in the time interval $\sim$\,$[20, 30]$\,min (top left panel) and the inferred chromospheric density at core height of $\lambda8542$, the wave energy flux is approximately 8.1\,$\rm kW\,m^{-2}$. This value is comparable to the perturbation in the radiative losses ($\sim$7.9\,$\rm kW\,m^{-2}$) in the same time interval (bottom left panel), well within the uncertainties. For reference, the time averaged deposited acoustic flux is only $Q_{\rm low}$\,$\sim$\,2.3$\rm\,kW\,m^{-2}$ at location (A). This is suggestive of heating by a shock wave in this particular time interval. 
Nevertheless, velocity oscillations in $\lambda8542$ as strong as the above are uncommon in the FOV, with only 10\% of the pixels showing $\delta v_{\rm LOS}$\,$>$\,$4\rm~km\,s^{-1}$. 

Locations (B) and (C) do not exhibit a clear modulation of the radiative losses in phase with the velocity and temperature oscillations. However, we do observe a slight increase in radiative losses above the mean values when the strongest velocity perturbations develop, approximately between $\sim$\,$[12, 20]$\,min in both cases.


\subsection{Inversions of IBIS+IRIS spectra: the plage region}
\label{section:Results:InversionsOfIBISandIRIS}

Figure~\ref{fig:inv2} presents an overview of the inversion results for the joint IBIS and IRIS observations of the small plage region and surroundings, as shown in Fig.~\ref{fig:FOV_overview}. Although the NNs offer good initial guesses for the NLTE inversions, inverting the \ion{Mg}{II} lines together with the visible lines for a long time series of a moderate size FOV remains prohibitively expensive. Therefore, we only inverted about $\sim$\,13.2\,min of data starting at 15:54\,UT. Furthermore, due to the IRIS lower sampling rate ($\sim$\,26\,s) compared to IBIS, we can only resolve frequencies up to $\sim$\,20\,mHz, making a detailed time-domain analysis unwarranted in this case.

While we obtained excellent fits to the IBIS lines alone, reconciling the intensities in the wings of $\lambda8542$ with the \ion{Mg}{II} h and k intensities proved challenging. The inversions typically underestimate the former by an average of 4\% compared to the observations (Fig.~\ref{fig:inv2}\textcolor{xlinkcolor}{b}). However, the $\lambda5896$ line and the range containing the \ion{Mg}{II} h, k, and UV triplet lines are well reproduced. Increasing the fitting weights of $\lambda8542$ relative to h and k did not improve the fits to $\lambda8542$ and only worsened the fits to the IRIS passband. Adding extra nodes at column masses in the photosphere/low-chromosphere did not significantly improve the quality of the fits. Although we expected that including the IRIS data in the inversions would reduce the bias in determining thermodynamic parameters, it also increased their variance, particularly raising the noise level in the $v_{\rm LOS}$ power spectra at different heights. Nonetheless, the quality of the fits is satisfactory, warranting further consideration and analysis.

For reference, the average formation height for the core of the k line in the inversion models is at $\log \xi$\,$\sim$\,$-5.2$ equivalent to $z$\,$\sim$\,$2100(\pm 400)$\,km in hydrostatic equilibrium. The corresponding mean($\pm$ standard deviation) mass density is $\rho$\,$\sim$\,$5(\pm 1)$\,$\times$\,$10^{-11}\rm\, kg\,m^{-3}$, that is 400 times smaller than the density at the formation height of $\lambda8542$ (cf. \S\,\ref{section:Results:formationNaCa}).

\subsubsection{Temperatures, radiative losses, and velocities}
\label{section:Results:InversionsOfIBISandIRIS_tvlos}

The inversions reveal a significant difference between the inferred temperatures from the IBIS and IBIS+IRIS modes, well beyond the estimated uncertainties indicated by the error bars (Fig.~\ref{fig:inv2}\textcolor{xlinkcolor}{a}). As expected, the differences are the largest (few tens of percent) for layers above the formation of $\lambda8542$, which are not constrained in the IBIS mode. Additionally, the temperature differences are also significant around the temperature minimum region, with the IBIS+IRIS inversions often showing a shift of the temperature minimum to a deeper column mass compared to the IBIS mode. Consequently, there is an increase in the integrated radiative losses in the low chromosphere by approximately $\sim$\,40\% on average (Fig.~\ref{fig:inv2}\textcolor{xlinkcolor}{d}) according to our criterion (\S\,\ref{section:methods_radlosses}), which is above the estimated uncertainties (\S\,\ref{section:methods_uncertainties}). However, the increase in mass density at the temperature minimum is disproportionately large (more than threefold on average), which will have a significant impact on the acoustic flux there. 

Analysis of the temperature response functions ($\partial I_\lambda/\partial T(\log \xi)$) reveals a gap in the spectral line sensitivity to temperature perturbations around the temperature minimum region ($\log \xi$\,$\sim$\,$[-2, -1]$) in both $\lambda5896$ and $\lambda8542$ lines (Fig.~\ref{fig:inv2}\textcolor{xlinkcolor}{b}). 
However, the \ion{Mg}{II} k (and h, not displayed) line show a more consistent and smoothly varying response function across that region (Fig.~\ref{fig:inv2}\textcolor{xlinkcolor}{c}), which theoretically helps to constrain the response of the visible lines (Fig.~\ref{fig:inv2}\textcolor{xlinkcolor}{b}). In practice, however, any unknown systematic effects in the data (calibration uncertainties, resolution, time cadence, etc.) or modeling (accuracy/completeness of atomic data, inversion setup, etc.) can make it difficult to simultaneously reproduce all the observed intensities. 

We have therefore chosen to evaluate the deposited acoustic and radiative losses only between the formation heights of the spectral line cores in the subsequent analysis of the plage target, as we cannot rule out systematic errors in the IBIS+IRIS inversion models around the temperature minimum region. Specifically, we only evaluate the acoustic fluxes and losses above the formation height of $\lambda5896$, where the temperature, density, and velocity distributions are more robustly determined in both IBIS and IBIS+IRIS models. For completeness, we present the results for both inversion modes. We note that the formation heights of the lines were carefully extracted for each pixel in the FOV and time frame rather than using averaged values. 

Similarly to the quiet patch, we verified that the PSDs from the IBIS+IRIS models also show broad peaks around $\sim$\,3\,$-$\,5\,mHz in the $\lambda5896$ and $\lambda8542$ lines, with increasing phase angles in the $\sim$\,5\,$-$\,12\,mHz range (not displayed). However, the PSD for $\lambda2796$ does not show significant power peaks, and the phase angle relative to $\lambda8542$ is approximately constant in frequency, which could be attributed to resonances and phase mixing at the top of the chromosphere.

\subsubsection{Relationship between losses and the magnetic field }
\label{section:BHinode}

\begin{figure}
    \centering
    \includegraphics[width=\linewidth]{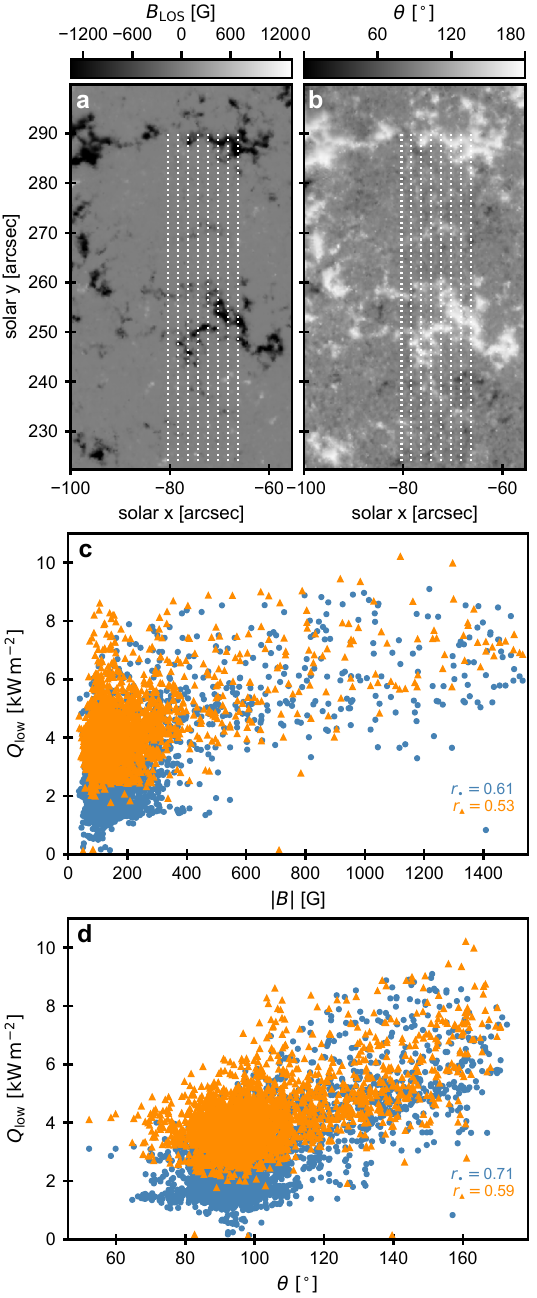}
    \caption{Photospheric magnetic field and radiative losses in the chromosphere. Top panels: closest-in-time Hinode LOS magnetogram (a) and inclination angle (b) corresponding to the IRIS/IBIS time series; dotted lines indicate the IRIS slit positions. Bottom panels: Radiative losses vs. total field strength (c) and inclination angle (d) for the IBIS (blue) and IBIS+IRIS (orange) inversions.} 
    \label{fig:hin}
\end{figure}

Figure \ref{fig:hin} displays the closest-in-time magnetic field strength and inclination maps in the photosphere provided by Hinode, compared to the time-averaged radiative losses in the low chromosphere for the IBIS and IBIS+IRIS inversion models. The magnetogram reveals an essentially unipolar plage region with a maximum field strength of approximately 1.5\,kG within the IRIS raster scan.

While there is a general spatial association between magnetic elements and enhanced radiative losses, the linear correlation between field strength and radiative losses is weak in both inversion modes. In contrast, the correlation between field inclination and radiative losses is stronger, with a correlation coefficient of $r$\,$=$\,0.71 in the IBIS mode, despite considerable scatter. More inclined fields may facilitate wave propagation and shock formation, leading to greater energy deposition and higher radiative losses.

The scatter can be attributed in part to the low cadence of the Hinode observations ($\sim$10\,min) compared to the IRIS rasters scans ($\sim$26\,s), as well as the inherent uncertainties of the radiative losses (\S\,\ref{section:methods_uncertainties} and \ref{section:muram_results}). Additionally, photospheric magnetograms may not be appropriate for detailed comparison with the chromospheric losses. We find that the correlation coefficients are higher for the IBIS mode than the IBIS-IRIS mode, which is attributable to higher inversion noise in the latter. We also investigated the differences between the velocities PSDs in strongly- and weakly-magnetized regions and found no significant differences.

\subsubsection{Deposited acoustic flux vs radiative losses}
\label{section:IRISdepflux}

\begin{figure}
    \centering
    \includegraphics[width=\linewidth]{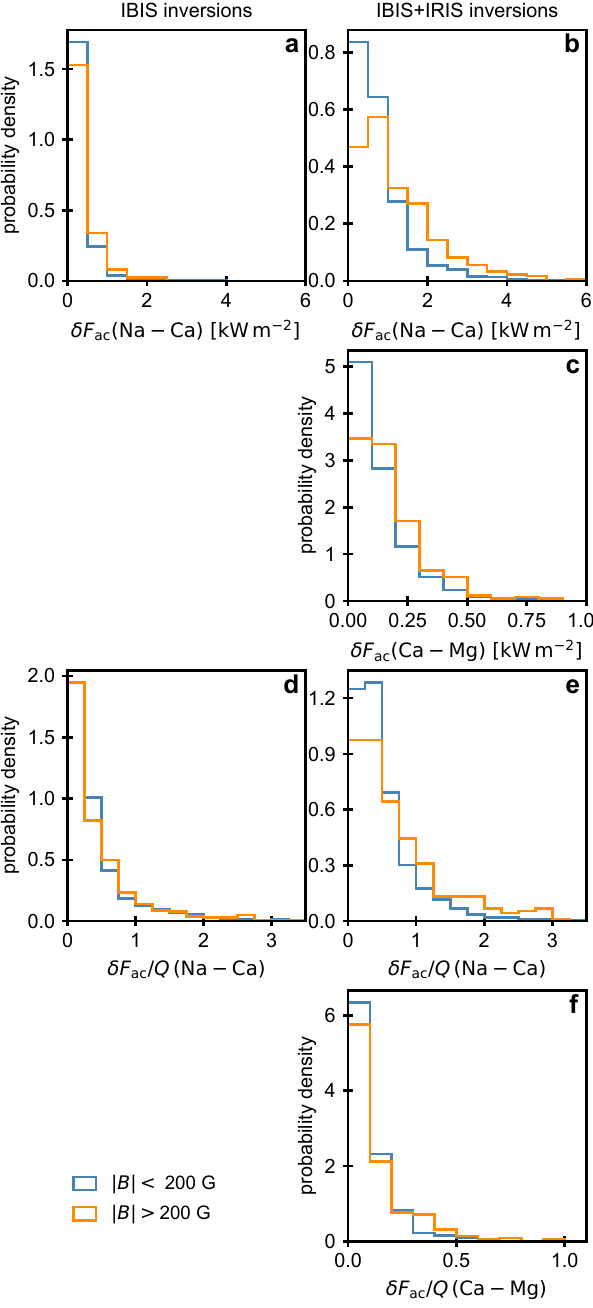}
    \caption{Deposited acoustic flux vs losses in the plage region. The left (right) columns show the results for the IBIS (IBIS+IRIS) inversions, separated for weakly (blue) and strongly (orange) magnetized regions. Panels (a-c): Deposited acoustic flux betweent the formation heights of sodium, calcium, and magnesium. Panels (d-f): The respective ratios of acoustic fluxes to the radiative losses.} 
    \label{fig:IRISFacQ}
\end{figure}

Figure \ref{fig:IRISFacQ} presents the histograms of deposited acoustic flux between the formation heights of $\lambda5896$ and $\lambda8542$, as well as between $\lambda8542$ and $\lambda2796$ for the plage region and surroundings. The figure also displays the ratio of acoustic flux to the integrated radiative losses. Similar to the quiet patch, the acoustic flux was integrated between the acoustic cutoff frequency and 20\,mHz. We accounted for the spatial variation of temperature (obtained from the NTLE inversions, \S\,\ref{section:Results:InversionsOfIBISandIRIS_tvlos}) and magnetic field inclination (provided by the Hinode inversions, \S\,\ref{section:BHinode}) when determining the cutoff frequency at each location in the FOV, assuming it remains constant within the Hinode raster time interval. However, we imposed a field strength limit of 150\,G to ensure the reliability of the inclination values.
To compare conditions in the weakly and strongly magnetized regions in and around the plage region, we split the analysis based on a magnetic field strength threshold of 200\,G, which effectively delimits the boundaries of the plage region (Fig.~\ref{fig:hin}). 

As expected, we find significant differences in the deposited acoustic flux between the core heights of sodium and calcium from both inversion modes, with the IBIS+IRIS mode yielding values over three times larger on average across the entire FOV (Fig.~\ref{fig:IRISFacQ}\textcolor{xlinkcolor}{a-b}). This is due to higher mass densities, velocity power, or a combination of both, which we need to interpret with caution. We excluded bad fits from the analysis based on an arbitrary $\chi^2$ (Eq.~\ref{eq:reg_chisq}) threshold (4$\sigma$), but it is still possible that spurious values, particularly in the IBIS+IRIS inversions, may affect the results.  
In the IBIS+IRIS inversions, distributions of deposited flux show a lower spread  (relative to the mean) and kurtosis between the formation heights of $\lambda8542$ and $\lambda2796$ than between $\lambda5896$ and $\lambda8542$, where 90\% of the time the values are below 0.3\,$\rm kW\,m^{-2}$ and 2\,$\rm kW\,m^{-2}$, respectively (Fig.~\ref{fig:IRISFacQ}\textcolor{xlinkcolor}{b-c}). This could be due to higher inversion noise deeper in the atmosphere or a real spatial uniformity of the acoustic fluxes in higher layers of the chromosphere.

The distributions of deposited acoustic flux are more right-skewed for the strongly magnetized regions at all sampled heights. However, with the exception of panel (b), the median values are identical for both regions. We obtained median values of $\delta F_{\rm ac}(\rm Na-Ca)$\,$\sim$\,0.2\,$\rm kW\,m^{-2}$ for the IBIS inversions (Fig.~\ref{fig:IRISFacQ}\textcolor{xlinkcolor}{a}) and $\delta F_{\rm ac}(\rm Na-Ca)$\,$\sim$\,0.6(1)\,$\rm kW\,m^{-2}$ for the IBIS+IRIS inversions in the weakly(strongly) magnetized regions (Fig.~\ref{fig:IRISFacQ}\textcolor{xlinkcolor}{b}). Between the formation heights of $\lambda8542$ and $\lambda2796$, we obtained $\delta F_{\rm ac}(\rm Ca-Mg)$\,$\sim$\,0.1\,$\rm kW\,m^{-2}$ for both regions (Fig.~\ref{fig:IRISFacQ}\textcolor{xlinkcolor}{c}). 

Similar to the quiet patch (\S\,\ref{section:Results:IBISdepflux}), the ratios of acoustic fluxes to the radiative losses are below unity for the majority of the pixels (Fig.~\ref{fig:IRISFacQ}\textcolor{xlinkcolor}{d-e}). Locations where the ratio surpasses unity are uncommon, corresponding to less than 10\% of the whole FOV for both inversion modes. These locations appear scattered around the magnetic elements, avoiding the strongest field concentrations, and some may be spurious results.
We obtained median values of $\delta F_{\rm ac}/Q(\rm Na-Ca)$\,$\sim$\,0.3  (Fig.~\ref{fig:IRISFacQ}\textcolor{xlinkcolor}{d}) for the IBIS inversions in both regions and $\delta F_{\rm ac}/Q(\rm Na-Ca)$\,$\sim$\,0.4(0.5) in the weakly(strongly) magnetized regions for the IBIS+IRIS models (Fig.~\ref{fig:IRISFacQ}\textcolor{xlinkcolor}{d-e}). 

Between the core heights of $\lambda8542$ and $\lambda2796$, we obtained $\delta F_{\rm ac}/Q(\rm Ca-Mg)$\,$\sim$\,0.07 for both regions (Fig.~\ref{fig:IRISFacQ}\textcolor{xlinkcolor}{f}), and the distributions shows a lower percentage of outliers than in the deeper layer; about 95\%(86\%) of the pixels show a $\delta F_{\rm ac}/Q(\rm Ca-Mg)$ ratio smaller than 30\% in the weakly(strongly) magnetized regions.

\subsection{Systematic errors in the velocities and cooling rates} 
\label{section:muram_results}

To assess the accuracy of the ratio $\delta F_{\rm ac}/Q$, we estimated the bias in the frequency-integrated acoustic flux rather than estimating an attenuation coefficient as a function of frequency (Eq.~\ref{eqn:acoustic_flux}), which is more susceptible to noise. This was done by comparing the velocity PSDs obtained from inversions of synthetic spectra from a \texttt{MURaM} simulation (\S\,\ref{section:methods_simulation}) with the true vertical velocities in the simulation at different heights. We also examined the accuracy of NLTE inversions in retrieving the radiative losses from the simulation.

The spectra were synthesized with \texttt{STiC} using the same atomic models, wavelength ranges and spectral sampling as in the inversions of the observations. We convolved the synthetic profiles with the instrumental spectral point-spread-functions, but we did not investigate the effect of spatial resolution and residual straylight. Due to the high computational cost of fitting the \ion{Mg}{II} h and k lines together with the visible lines, we only synthesized and inverted one (of 100) time stamps including the IRIS lines for comparing the radiative losses of the IBIS and IBIS+IRIS models. Consequently, we are unable to compare the simulated and inverted velocity PSDs in $\lambda2796$. Similar to the observations, a NN could, in principle, accelerate the inversions. However, this would require its own separate training and validation processes, which is beyond the scope of this paper. Nevertheless, we expect the attenuation coefficient in $\lambda2796$ to be slightly smaller than in $\lambda$8542 \citep{2023arXiv230204253M}. 

\begin{figure}
    \centering
    \includegraphics[width=\linewidth]{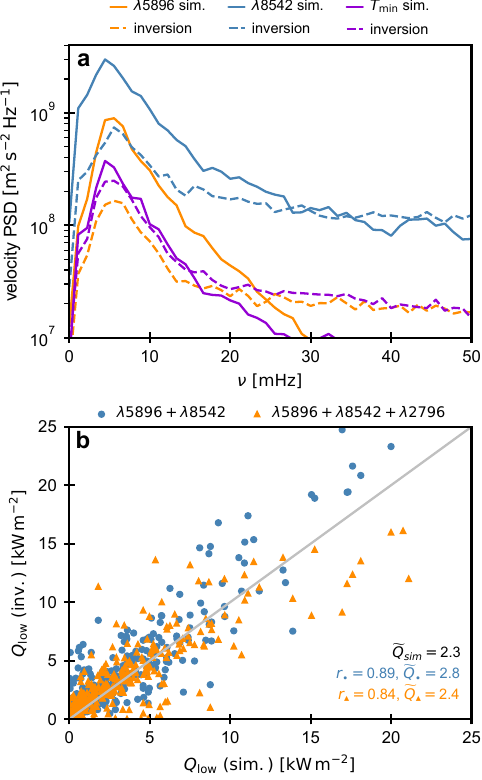}
    \caption{Velocity power spectra and radiative losses in a \texttt{MURaM} simulation. Panel (a): LOS velocity power spectral density in the simulation (solid lines) and the inverted models (dashed lines) at the formation heights of $\lambda5896$ and $\lambda8542$ and at the temperature minimum height as a function of frequency. Panel (b): Simulation vs inversion total cooling rates in the low chromosphere for the IBIS and IBIS+IRIS models; the gray line shows the 1:1 locus; correlation coefficients and median values are displayed on the lower right.} \label{fig:muram}
\end{figure}

Figure \ref{fig:muram} presents the true and inverted velocity PSDs and radiative losses. The inversions successfully recover the dominant velocity power peak at all heights. However, beyond 20\,mHz, the PSDs flatten due to noise, and the inferred PSDs fall below the ground truth. The agreement is better at the temperature minimum height and worsens at higher heights. On average, the recovered acoustic flux between 5\,$-$\,20\,mHz is 70\% at the temperature minimum but drops to 20\% at the $\lambda$8542 core height. Despite this, because the acoustic flux is much lower at higher heights, the deposited acoustic flux between these two heights remains fairly robust. 

We also find that both the IBIS and IBIS+IRIS inversions reproduce the simulated radiative losses reasonably well, with strong positive correlations ($r>0.8$) between the inferred and simulated values. Although the IBIS+IRIS mode shows a slightly smaller $r$ value than the IBIS mode, the scatter is reduced in the former, especially for losses below 5\,$\rm kW\,m^{-2}$, and the median value of the distribution ($\widetilde Q$\,$\sim$\,2.4\,$\rm kW\,m^{-2}$) is more similar to the ground truth ($\widetilde Q$\,$\sim$\,2.3\,$\rm kW\,m^{-2}$). However, the IBIS-only models tend to overestimate the radiative losses ($\widetilde Q$\,$\sim$\,2.8\,$\rm kW\,m^{-2}$) compared to the IBIS+IRIS models, whereas in the observational data, we find that the IBIS+IRIS inversions yield larger values (\S\, \ref{section:IRISdepflux}). The larger difference between the two inversion modes in the observations compared to the simulation could be due to systematic errors in the observational data or differences in solar conditions from those simulated. These statistics are likely model-dependent. In higher layers of the atmosphere, between the formation heights of $\lambda8542$ and $\lambda2796$, the losses are relatively better constrained, typically being overestimated by 14\%.


\section{Discussion and Conclusions}

In this paper, we quantified the ratio of the deposited acoustic flux to the radiative losses at different heights under quiet and active conditions using multi-atom, multi-line, NLTE inversions. This technique enabled us to obtain self-consistent velocity power spectra and cooling rates at each location in the FOV. This approach is reminiscent of the grid-based methods employed by \citet{2016ApJ...826...49S} and \citet{2020A&A...642A..52A,2021A&A...648A..28A}, but differs in that we did not use a precomputed grid of temperature and density models external to the data. Instead, we inferred models from the observations that additionally included the LOS velocities for each time frame and pixel in the FOV. Moreover, we compared the wave fluxes and losses in somewhat deeper chromospheric layers than the previous studies, thanks to the $\lambda5896$ observations. 

\citet{2023A&A...670A.133F} showed that, although inversions provide a hydrostatic model for a dynamic atmosphere, they still yield meaningful atmospheric parameters in the presence of oscillations, reliably recovering dominant wave power peaks up to at least 9\,mHz in their sunspot umbra simulation. This shows that inversions are reasonable tools for determining local thermodynamic variables and quantifying wave fluxes  \citep[see also][]{2021RSPTA.37900182K}. We have further explored this approach through inversion experiments on synthetic data from a QS simulation performed with the \texttt{MURaM} code, allowing us to investigate systematic effects in the inferred velocity PSDs at different heights. 

Our results indicate that NLTE inversions may significantly underestimate acoustic fluxes in higher layers ($\sim$80\%) of the chromosphere compared to the base ($\sim$30\%) while simultaneously slightly overestimating the radiative losses. The bias is stronger when the spectral diagnostics are more limited (i.e., only $\lambda5896$ and $\lambda8542$), as the variations of temperature, density and velocity with height are not perfectly constrained. This suggests a slight increase in the contribution of waves to the total cooling rates in the low chromosphere. Nevertheless, even accounting for this bias, the ratio of acoustic flux to radiative losses remains below unity at most locations within the quiet patch and plage region. These estimates may vary depending on the inversion setups, particularly when using different combinations of spectral lines than those employed in this study.

We note that the (velocity) attenuation effects of the atmosphere are model-dependent \citep[e.g.,][]{2023arXiv230204253M}. Therefore, this numerical experiment should be regarded as illustrative. To draw more general conclusions, it would be necessary to repeat the experiment across a range of simulated solar-like conditions to investigate the response of spectral lines to wave propagation and dissipation in greater detail. Specifically, we did not explore the impact of magnetic fields on the attenuation coefficient and losses. Nonetheless, we expect the coefficient to be close to one above magnetic field concentrations \citep{2023arXiv230204253M}, meaning that this parameter is less critical when interpreting the acoustic fluxes in plage regions compared to the internetwork. 

The inversions of the IBIS-only dataset provided excellent fits to the visible lines, enabling the reconstruction of the lower chromosphere's stratification for both the quiet patch and the plage region. The inclusion of IRIS data for part of the FOV offers the potential to better constrain the velocity and temperature power spectra as a function of height in the plage region, diagnosing higher layers than possible with IBIS alone. We anticipated that the \ion{Mg}{II} h and k lines would significantly impact the determination of the thermodynamic parameters of the atmosphere \citep[e.g.,][]{2016ApJ...830L..30D, 2018A&A...620A.124D, 2023A&A...672A..89K}. We confirm that this expectation translates into more accurate radiative losses in the low chromosphere.
However, fitting the $\lambda5896$, $\lambda8542$, and \ion{Mg}{II} h and k lines simultaneously proved more challenging than fitting the visible lines alone. This resulted in discrepancies in the core of $\lambda5896$ and, especially, in the far wings of $\lambda8542$. These issues arose from the inversions' difficulty in producing a temperature/density stratification around the temperature minimum that could simultaneously reproduce the observed intensities.  
Given the suboptimal goodness of the fit of the IBIS+IRIS inversions, we must interpret the differences between the IBIS and IBIS+IRIS models with caution.  

We note that inversions of IBIS and IRIS data have not previously been attempted. The observed fitting residuals might be due to an unknown offset in the IRIS absolute flux and/or wavelength calibrations, inaccurate modeling of the (spectral) point-spread functions of the IRIS and/or IBIS instruments, or the lower spatial resolution of the IRIS data compared to the IBIS data, which might necessitate a more detailed treatment of instrumental degradation effects. We investigated whether adjusting the UV intensities by a few percent improved the inversions, given their larger relative uncertainty, but the results were inconclusive. Modifying the number of nodes and regularization weights also did not resolve the misfits. However, the fact that \citet{2023A&A...672A..89K} were able to fit $\lambda$8542 (observed at the Swedish Solar Telescope) and $\lambda2796$ using the same atomic data, despite similar spatial and temporal resolution constraints, suggests that issues with our inversion setup or the IRIS data are less likely. We could not rule out potential calibration problems with the IBIS data. 

Even though we find a clear positive phase differences between the $\lambda$5896 and $\lambda$8542, indicating upward-propagating waves, the estimated acoustic wave energy flux is insufficient to balance the radiative losses in the low chromosphere ($\sim$2.8\,$\rm kW\,m^{-2}$) in the quiet patch, at least for frequencies below 20\,mHz. However, the wave flux is not negligible; on average, it accounts for about one-third of the losses in the quiet patch, or about half if we took the effects of power spectra attenuation and the overestimation of losses inferred from the \texttt{MURaM} simulation at face value. This places our estimate within the range of values (spanning an order of magnitude) reported in the literature \citep[e.g.,][]{2005Natur.435..919F,2006ApJ...646..579F,2007PASJ...59S.663C,2016ApJ...826...49S,2020A&A...642A..52A, 2021A&A...648A..28A, 2021ApJ...920..125M}. 

We assumed the acoustic waves to be exclusively vertically propagating \citep{2006ApJ...646..579F}, though obliquely propagating waves could increase the total energy flux by a small amount \citep{2010A&A...522A..31B}. On the other hand, we did not include the contribution of the \ion{Fe}{II} lines to the cooling rates \citep{1989ApJ...346.1010A}, as we lacked a suitable atomic model, so our radiative losses should be regarded as lower limits.

Higher up in the chromosphere in the weak plage region, between the formation heights of $\lambda8542$ and $\lambda2796$, the acoustic fluxes are clearly insufficient to explain the observed emissions, contributing to less than 10\% to the radiative losses. Our estimate is on the lower end of the range of values found in the recent literature \citep[10$-$30\%,][]{2021A&A...648A..28A, 2022A&A...664A...8M}. In this case, we expect the radiative losses to be relatively well constrained based on the simulation results; as such, even if the acoustic flux suffered a reasonable attenuation \citep{2023arXiv230204253M}, it still would not balance the losses. 
Moreover, the distributions of the radiative losses are narrower and fairly uniform in space, with only a slight skew towards higher values above strong field concentrations. This suggests a more uniform energy deposition in higher chromospheric layers, potentially due to a more uniform magnetic field there \citep{2021SciA....7.8406I}. 

Nevertheless, we observe instances where the ratio of deposited acoustic flux to radiative losses significantly varies over time, approaching unity when strong velocity/temperature perturbations develop, leading to increased cooling rates. These perturbations, likely linked to shocks, suggest that wave energy contributes effectively to localized chromospheric heating. In contrast, some areas show only modest increases in losses associated with waves, indicating that the relationship between dynamics and radiative losses is complex and can vary widely across different regions. 

Based on NLTE inversions of the \ion{Ca}{II} K line, \citet{2022A&A...668A.153M} reported magnitudes of velocity and temperature perturbations in bright grains that are similar to our inversion results from $\lambda8542$ in the quiet patch. However, while the authors found temperature enhancements associated to upflows, we find them in tandem with downflows, at least at the formation height of $\lambda8542$. At the same time, temperatures decrease at the core height of $\lambda5896$. We speculate that the $\lambda8542$ line is not as sensitive to the hot shock front propagating upward as the K line but only detects the return flows while the atmosphere is still being heated. Conversely, $\lambda5896$ probes the plasma beneath the shock, as it cools down through adiabatic expansion. Simulations may also shed light on these processes.

The remaining energy budget in both plage regions and the internetwork could be explained by additional processes mediated by the magnetic field, possibly involving small-scale magnetic reconnection \citep[e.g.,][]{2018ApJ...857...48G,2024ApJ...964..175G, 2019ApJ...878...40M,2020A&A...641L...5J}, Alfvén waves \citep[e.g.,][]{1961ApJ...134..347O,2001ApJ...558..859D,2013ApJ...777...53T,2023ApJ...958...81M}, and ion-neutral effects \citep[e.g.,][]{2005A&A...442.1091L,2012ApJ...747...87K,2016ApJ...817...94A,2017ApJ...840...20S,2020A&A...635A..28W, 2021A&A...652A.124N}. The observed correlation between the photospheric field inclination and the chromospheric radiative losses hints to the significance of the magnetic field topology in the heating processes, influencing wave propagation and dissipation mechanisms \citep[e.g.,][]{2006ApJ...648L.151J,2009A&A...494..269V,2010A&A...510A..41K}. 

Our analysis highlights the importance of considering spatially resolved data when studying the chromospheric energy balance. The significant variability observed in both wave fluxes and radiative losses underscores the need for high-resolution observations to accurately capture the dynamics of the solar atmosphere. 
Future work should incorporate multi-height spectropolarimetry, which is currently lacking in our analysis but can be easily accommodated by existing NLTE inversion codes. In this regard, the Daniel K. Inouye Solar Telescope \citep[DKIST,][]{2020SoPh..295..172R} is well-suited to advance this research through multi-line spectropolarimetry from the photosphere to the chromosphere. Coordination with IRIS will offer complementary wavelength coverage, providing intensities in the \ion{Mg}{II} h and k lines and enabling tighter constraints on radiative-transfer-based models.

\begin{acknowledgments}
The NSO is operated by the Association of Universities for Research in Astronomy, Inc., under cooperative agreement with the National Science Foundation. IBIS has been designed and constructed by the INAF/Osservatorio Astrofisico di Arcetri with contributions from the Università di Firenze, the Università di Roma Tor Vergata, and upgraded with further contributions from NSO and Queens University Belfast. This work utilized the Blanca condo computing resource and the Alpine high performance computing resource at the University of Colorado Boulder. 
Blanca is jointly funded by computing users and the University of Colorado Boulder. Alpine is jointly funded by the University of Colorado Boulder, the University of Colorado Anschutz, and Colorado State University.
The Institute for Solar Physics is supported by a grant for research infrastructures of national importance from the Swedish Research Council (registration number 2021-00169). 
JdlCR gratefully acknowledges financial support by the European Union through the European Research Council (ERC) under the Horizon Europe program (MAGHEAT, grant agreement 101088184). 
IM acknowledges the funding provided by the Ministry of Science, Technological Development and Innovation of the Republic of Serbia through the contract 451-03-66/2024-03/200104.
\end{acknowledgments}

\vspace{5mm}
\facilities{SDO(HMI), DST(IBIS), IRIS, Hinode(SOT)}

\software{ \texttt{Astropy} \citep{2013A&A...558A..33A,2022ApJ...935..167A},
\texttt{Sunpy} \citep{sunpy_community2020},
\texttt{CMasher} \citep{2020JOSS....5.2004V}, \texttt{PyTorch} \citep{NEURIPS2019_9015},
          \texttt{STiC} \citep{2016ApJ...830L..30D,2019A&A...623A..74D}
          }



\appendix

\section{Neural-network-assisted inversions}
\label{section:appendix}

\begin{figure}
    \centering
    \includegraphics[width=0.4\linewidth]{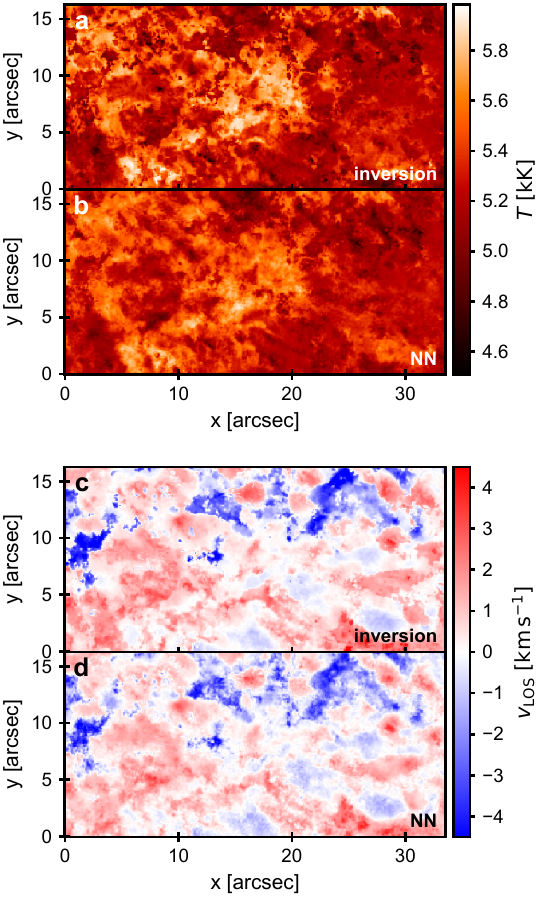}
    \caption{Neural network predictions vs inversions. Temperature (top panels) and LOS velocity (bottom panels) at a column mass in the chromosphere ($\log \xi$\,$\sim$\,$-3$.) obtained through a full inversion ((a) and (c)) and the NN prediction ((b) and (d)) of the quiet patch (cf. Fig.\,\ref{fig:FOV_overview}).} \label{fig:NN} 
\end{figure}

Inversions of multiple NLTE spectral lines, particularly those requiring PRD calculations, are computationally intensive. This is primarily due to the need to solve the RT equation multiple times, in various propagation directions, to obtain a self-consistent solution for the statistical equilibrium equation governing the number densities of the atomic states involved in the line formation \citep[e.g.,][]{1991A&A...245..171R}. Consequently, every evaluation of the chi-squared value (Eq.\,\ref{eq:reg_chisq}) requires hundreds of RT iterations. 
Inverting a single spectrum with \texttt{STiC} can therefore take from several minutes to a few hours of CPU time, making the inversion of time-dependent, high-resolution observations prohibitively expensive. 

One approach to expedite inversions is to ensure that the initial guess of the optimization algorithm closely approximates the best fit. To achieve this, we employed Neural Network-assisted inversions. Neural networks (NNs) are machine learning methods that employ a series of simple, non-linear transformations. These transformations are optimized to enable the network to "learn" how to produce an output based on the input training set, which, in this case, involves mapping observed intensities to atmospheric parameters \citep[e.g.,][]{2019A&A...626A.102A}.

We trained a simple, fully-connected NN using \texttt{PyTorch} \citep{NEURIPS2019_9015} to map the input spectra to the values of the relevant physical parameters at the set of nodes \citep[see also][]{2021A&A...651A..31G}. For the training set, we inverted four data cubes from the IBIS time series, three of which were used for training and validation (178\,500 spectra) and one for testing (i.e., three cubes were used to constrain the parameters of the network and one to verify the network accuracy). The input to the NN consisted of the normalized spectral line intensities at different wavelengths, while the output contained the values of $T$, $v_{\rm turb}$, and $v_{\rm LOS}$ at the set of the nodes used for the inversion (\S\,\ref{section:methods_nlteinversions}). 
The training process took several hours on a desktop GPU (\texttt{Nvidia GeForce 3060 Ti}). In turn, the NN parameter inference from the input spectra only took milliseconds per spectrum. 

Testing showed that the NN predictions were very close to regular inversions, especially in $v_{\rm LOS}$ (Fig.\,\ref{fig:NN}). Still, to ensure fully self-consistent inversions, we did not use the NN output as the final results for the analysis but as initial solutions for regular \texttt{STiC} inversions. This resulted in an effective speedup of about a factor of four. In the future, we will investigate whether the NN can also directly retrieve radiative losses, thereby eliminating the need for an additional round of RT synthesis to derive them from the inversion models.


\bibliographystyle{aasjournal}

\end{document}